\documentclass[smallextended]{svjour3}       % onecolumn (second format)

\smartqed  % flush right qed marks, e.g. at end of proof

\usepackage{amsmath}
\usepackage{amssymb}
\usepackage{dsfont}
\usepackage{listings}
\usepackage[utf8]{inputenc}
\usepackage{paper_alcides}
\usepackage{algorithm}
\usepackage{algpseudocode}
\usepackage{url}
\usepackage{subfigure}

\journalname{International Journal of Parallel Programming}
\begin{document}

\title{Automatic Parallelization: Executing Sequential Programs on a Task-Based Parallel Runtime
\thanks{This work was partially supported by the Portuguese Research Agency FCT, through CISUC (R\&D Unit 326/97) and the \texttt{CMU|Portugal} program (R\&D Project Æminium CMU-PT/SE/0038/2008). }}


\author{Alcides Fonseca\thanks{Supported by the Portuguese National Foundation for Science and Technology (FCT) through a Doctoral Grant (SFRH/BD/84448/2012).}     
\and Bruno Cabral
\and João Rafael
\and Ivo Correia
}

\institute{Alcides Fonseca, Bruno Cabral, João Rafael, Ivo Correia \at
              Department of Informatics Engineering \\
              Universidade de Coimbra \\
              Tel: +351 239 790 000 \\
              Fax: +351 239 701 266 \\
              \email{\{amaf, bcabral\}@dei.uc.pt}, \\
              \email{\{jprafael, icorreia\}@student.dei.uc.pt}
}

\date{Received: date / Accepted: date}

\maketitle

\begin{abstract}

There are billions of lines of sequential code inside nowadays’ software which do not benefit from the parallelism available in modern multicore architectures. Automatically parallelizing sequential code, to promote an efficient use of the available parallelism, has been a research goal for some time now.

This work proposes a new approach for achieving such goal. We created a new parallelizing compiler that analyses the read and write instructions, and control-flow modifications in programs to identify a set of dependencies between the instructions in the program. Afterwards, the compiler, based on the generated dependencies graph, rewrites and organizes the program in a task-oriented structure. Parallel tasks are composed by instructions that cannot be executed in parallel. A work-stealing-based parallel runtime is responsible for scheduling and managing the granularity of the generated tasks. Furthermore, a compile-time granularity control mechanism also avoids creating unnecessary data-structures.

This work focuses on the Java language, but the techniques are general enough to be applied to other programming languages.

We have evaluated our approach on 8 benchmark programs against OoOJava, achieving higher speedups. In some cases, values were close to those of a manual parallelization. The resulting parallel code also has the advantage of being readable and easily configured to improve further its performance manually.

\end{abstract}

\section{Introduction}

Nowadays, in order to achieve the best performance on multicore machines, programmers have to write parallel programs. This is typically done using threads, either directly or indirectly through other high-level constructs of the language. Traditionally, manually defining threads and synchronizing them is the only way to achieve the best results. However this process is often cumbersome and error-prone, often leading to the occurrence of problems such as deadlocks and race conditions. Furthermore, as the code base increases it becomes increasingly harder to detect interferences between executing threads. Writing, debugging and tuning multi-threaded code is very time-consuming, as there are multiple combinations of executions that make the performance and visibility of errors non-deterministic. Furthermore, there are billions of lines of source code inside existent software that are not able to benefit from today’s multicore architectures. Parallelizing these programs is a daunting and extremely costly task, one that hardly someone is eager to initiate.

The automatic parallelization of existing software has been a long running objective and prominent research subject \cite{banerjee1993automatic}. Existing research has been mainly focused on the analysis and transformation of loops, since these have always been perceived as the main source of potential parallelism in sequential programs \cite{feautrier1996automatic}. Nonetheless, other models have also been studied, such as the parallelization of recursive methods \cite{bik1997automatically}, and of sub-expressions in functional languages. Focusing only on the parallelization of loops is not enough in most cases and, other approaches have not revealed significant performance improvements.

In this paper we introduce a new approach for performing a fine-grained automatic parallelization of programs. This approach is distinct from others, since it parallelizes all the instructions that can, effectively, be executed in parallel. To identify which instructions can be parallelized, we infer instruction signatures from the source code of the program. These signatures include dependency and control flow information, which allows us to organize instructions into a task-oriented structure. The result is a program that exhibits the maximum possible parallelism at the finest granularity level (e.g. one task can equal one instruction). However, in order to achieve good performance and decrease the overhead in run-time task management, the granularity of tasks is coarsened during compilation and also during run-time. At run-time, the system load influences granularity control. Furthermore, a work-stealing scheduler is used to efficiently manage execution and control dependencies.

Our approach can parallelize irregular recursive programs with a low runtime overhead, resulting on up to 20x of speedup, on a 24 thread machine and an average of 5x of speedup. Because of dependency tracking and transformations during compilation, we are able to avoid harsh runtime overheads from which existing solutions suffered. This paper contributes with an hybrid methodology for analyzing procedural source code and translating it to a parallel version with a broad level of parallelism and granularity, that is fine-tuned during execution by runtime granularity control mechanisms. The parallelization approach was tested with popular benchmark tests for task-based parallelism, and compared with another two approaches.

We have applied this approach to the Java language, one of the most popular programming languages, since it has a large code base of legacy sequential software. Nonetheless, our approach can be applied to any procedural or object-oriented language.

The framework presented includes two language front-ends, the Æminium language compiler \cite{stork2014aeminium}, the JPar compiler for Java, the Æminium Runtime, and a ÆminiumGPU\cite{fonseca2013aeminiumgpu} compiler and runtime, to enable GPU execution of data-parallel programs.

This paper is organized as follows: This section introduced the Æminium framework; In Section~\ref{sec:related_work} we present the related work in automatic parallelization; In Section~\ref{sec:architecture} we describe in further detail the architecture of the framework, mainly from the dynamic perspective; In Section~\ref{sec:compilation} we explain the parallelization technique applied; In Section~\ref{sec:runtime} we present the Runtime support for executing parallel programs. Finally in Section~\ref{sec:evaluation} we evaluate the platform in different programs; and in Section~\ref{sec:conclusion} we lay the conclusions.

\section{Related Work}
\label{sec:related_work}

Given the wide availability of multicore processors, GPUs and other accelerators such as FPGAs and the Xeon Phi, research on concurrent programming has increased in the last decade. New programming models, languages and runtime systems have been developed to improve the expression and execution of parallel programs. Much of this work has culminated in new languages, such as X10\cite{charles2005x10}, Fortress\cite{steele2006parallel} and Chapel\cite{chamberlain2007parallel}, in which most language constructs are default by parallel (such as for cycles, for instance). These languages also provide constructs to explicitly inform the compiler that certain memory regions are independent and, therefore, accesses to them can be executed in parallel. Unlike these languages, which mostly target scientific computing, the Æminium language\cite{stork2014aeminium} has focused on dependable systems programming. By annotating variables with access permissions, programs could be automatically parallelized with guarantees that the execution would not break the defined contracts.

Another approach for writing parallel programs is semi-automatic parallelization. In this approach, programmers annotate existing sequential programs with enough information for the compiler to automatically parallelize parts of the code. Cilk\cite{frigo1998implementation} and OpenMP\cite{dagum1998openmp} are the two most common examples of such approach and work on top of the C language. Cilk focus on divide-and-conquer recursive algorithms, while OpenMP has focus mostly on symmetrical parallelism in for cycles. OpenMP 3.0 has introduced unstructured parallelism via the concept of Tasks\cite{ayguade2009design}\cite{ayguade2008experimental}. More recently, OpenMP has also started to generate code for GPUs\cite{lee2009openmp}.

The third approach is to translate unmodified sequential programs automatically to a parallel version of themselves. Functional Languages, like LISP\cite{hogen1992automatic} and Haskell\cite{marlow2009runtime}, can be easily parallelized since sub-expressions do not interferes with each other. Imperative languages such as C and Java make this task more difficult, since different parts of the code can access the same memory location. In order to be able to parallelize code, there are some verifications that have to be made. The main focus of research has been the parallelization of for loops. Different techniques can be used, depending on the type of loop, such as \textit{DO-ALL}, \textit{DO-ACROSS} or \textit{DO-PIPE}. \textit{DO-ALL} parallelism does not contain any interference between loop iterations, and each iteration (or sets of iterations, called slices), can be executed in parallel and they must synchronize at the end of the for-cycle. \textit{DO-ACROSS} has a part of the cycle (usually minimal compared to the rest) that interferes with other iterations. For these cases, variable privatizing can be done to aggregate values per thread, and then another for-cycle is sequentially executed in the end to aggregate the private variables. \textit{DO-PIPE} can be parallelized by using different threads for different parts of the for cycle, in which dependencies between different threads are minimized. In order to verify if the loops can be parallelized or not, the Polyhedral Model is frequently used\cite{bondhugula2008automatic}. Cetus\cite{dave2009cetus} and Par4All\cite{amini2012par4all} are compilers that perform this kind of transformation, which can also target GPGPUs. Loop parallelization has been done during runtime\cite{zhao2005loop}, but without any relevant speedups.

Automatic non-loop parallelization has been less studied, but it is still a popular way of expression parallelism, specially in divide-and-conquer algorithms. This analysis has been implemented in zJava\cite{chan2004run} by analyzing data writes and reads at a local level. In order to allow a correct parallelization, zJava uses a runtime registry of regions, to which threads can be assigned. This allows threads to access shared data with a synchronization overhead. OoOJava\cite{jenista2011ooojava} also performs static analysis to retrieve read and write information on annotated tasks. Then it compiles the program to a speculative C program, that has runtime checks to resolve conflicts between threads. Because of such speculation, OoOJava does not support I/O instructions. MP-Tomasulo\cite{wang2013mp} also uses Out-of-Order instructions for automatically parallelizing code for FPGAs but removes write-after-read and read-after-write by renaming parameters, keep control flow separate. This approach is not compatible with regular multicore processors, however. Jrpm\cite{chen2003jrpm} also performs thread-level speculation at runtime, operating over Java bytecode instead of Java code, revealing a worse speedup than compiler-time strategies. FJComp\cite{senghor2013fjcomp} also focus on Divide-and-Conquer algorithms using the Fork-Join framework. However, the compiler requires the programmer to annotate tasks and optionally define the cut-off mechanism, making FJComp more of a Translator from recursive calls to FJ-style calls, than an automatic parallelizing compiler.

\section{Architecture}
\label{sec:architecture}

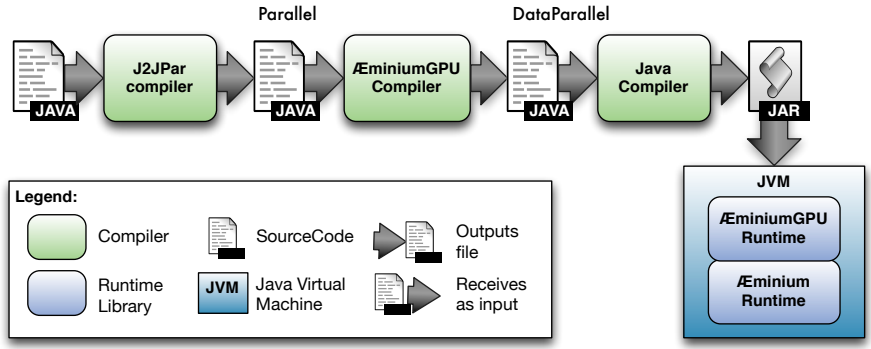
\botafig[0.5]{architecture_flow}{Information Flow in the Æminium Framework}

The software architecture of the Æminium framework is heavily based on the Java stack, making use of the Java Compiler to generate bytecode and a JVM for executing such bytecode. The framework includes two compilers:the JPar compiler and the ÆminiumGPU compiler. The later two make use of the Spoon library\cite{spoon} to operate on the Java Abstract Syntax Tree (AST). Figure~\ref{fig:architecture_flow} shows the information flow between the framework components. The ÆminiumGPU components are optional, as not all programs can take advantage of the GPU.

The JPar Compiler is a Source-to-Source Compiler for the Java language. It parses the original Java code into an AST, performs static analysis to infer access permissions of each node, and generates Java code that wraps some of the operations in calls to the Æminium Runtime.

The Æminium Runtime is a Java library that provides an API for expressing task-parallelism. These APIs can be targeted by compilers, or directly by programmers. The API allows the definition of tasks and dependencies between tasks. Tasks are wrappers around a set of Java statements that can execute asynchronously, and can have any number of instructions. Internally, the most important components are the Scheduler, Decider and Profiler. The Runtime includes a Work-Sharing scheduler, but defaults to a Work-Stealing scheduler, in which the programmer can configure the stealing policy. The Decider is a component that determines in real-time whether a new task should be created or if it should be inlined in the caller site. Finally, the Profiler records information during execution, such as number of tasks created, steals and dependencies unfulfilled.

The ÆminiumGPU Compiler is a Source-to-Source Compiler for the Java language that identifies data-parallel operations, such as map and reduce, and translates them to OpenCL. The ÆminiumGPU Runtime is a library that executes data-parallel operations on the GPU. During execution, if the operation is heavy and/or operates on a large dataset, the runtime decides whether to use the GPU or not. If it does, the JavaCL binding library\cite{dominguez2013evaluating} is used to schedule OpenCL code and copy memory between the host and the GPU.

\section{Compilation}
\label{sec:compilation}

This section details the compilation process, starting from access permission analysis to code generation. The JPar is based upon the J2JPar Compiler\cite{rafael2014dependency}, but simplifies the access permission analysis, and performs parallelization during code-generation, reducing compilation times. An overview of the compilation phase can be seen in Figure~\ref{fig:jpar}.

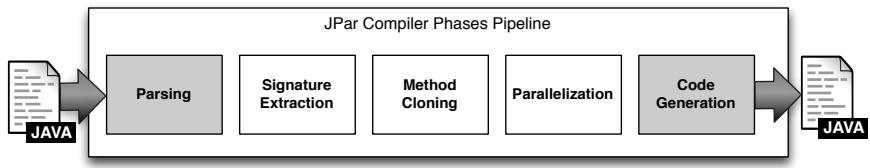
\botafig[0.51]{jpar}{JPar Compiler Phases}

\subsection{Signature Extraction}

In order to automatically parallelize the program, it is necessary to analyze the memory is accessed to understand dependencies between parts of the program. If two program parts read and write to the same variable, then they cannot be parallelized without guaranteeing determinism. Thus, the first step of the compiler is to understand what each AST node reads and modifies. Datagroups\cite{leino2002using} are used to represent different memory sections and if two method calls share no datagroup, it means that they can be executed in parallel. After this phase, each AST node will have a signature, composed of one or more datagroups permissions. An example of signatures in code can be seen in Listing~\ref{code:signatures}. Datagroup permissions can be one of the following: 

\begin{itemize}
	\item \textbf{read(dg)} - the AST subtree reads the variables represented by datagroup $dg$;
	\item \textbf{write(dg)}- the subtree writes to the objects in datagroup $dg$;
	\item \textbf{control(dg)} - the control flow of other operations in datagroup $dg$ may be altered. This is the case with return statements, breaks, continues, ifs and whiles.
\end{itemize}

In the previous approach, a full analysis of the AST was possible due to a two-pass verification. In the first pass, invocations produced a \textbf{call(dg)} permission, and aliasing effects produced \textbf{merge(dg1, dg2)} permissions. On the second pass, these permissions were replaced with the true permissions, that could be looked up in the rest of the program. The new approach converts the two passes into one. Aliasing is handled in-place, using HashMaps to find the right permission of the aliased element. When finding a method invocation, the current element being processed is saved, and the compiler processes the method declaration first. When it is complete, it returns to the method invocation and the method data-group is already available. The only place in which this is not possible is in recursive (direct or indirect) calls. In this case, the stack detects loops in the recursion, uses the partial permissions, to fill in the recursive invocation for the full permission. This process is now a two-step process only for recursive calls, reducing the analysis time in all other operations.

\lstset{language=Java,caption={Examples of Signatures in Fibonacci Program},label=code:signatures}
\begin{lstlisting}
int f(int n) {
    if (n < 2) { // read(n), control(f)
        return n; // write(return), control(f)
    }
    int a = f(n - 1); // call(f), read(n), write(a)
    int b = f(n - 2); // call(f), read(n), write(b)
    return a + b; // read(a), read(b), control(f), write(return)
}
\end{lstlisting}

In this phase, whenever some operations can have different results, such as the case of an \texttt{if} statement, a conservative approach is take, leaving the union of the two possible branches, as the signature for that node. This approach does not perform thread-level speculation, guaranteeing instead the same semantics of the original programming and supporting I/O and other operations that cannot be transactional. One such example is that all accesses to external objects, such as the System.out object, are inside a single global datagroup. This bottleneck can be removed by explicitly expressing the signatures for those methods in a special signatures file.

\subsection{Method Cloning}

Executing a method in parallel may not always be worthwhile. For instance, in the Fibonacci example, the cost of creating a new task is higher than the cost of executing the method for a low input number. Thus, creating a task is only useful when another thread can execute it. As such, the alternative is to execute the original method sequentially after a certain point. 

In this compiler phase, methods are cloned. The original method will be parallelized, while the clone will serve as a backup sequential version of the method. The decision when to change to the sequential method is introduced on the beginning of the parallel version. The decision itself is a call to the Runtime API.

For recursive calls, using anonymous inner classes revealed to be a very big overhead, which is not noticeable with regular parallel tasks. As such, in this phase, for each recursive call in the code, a static class is created to represent the asynchronous call to that method.

\subsection{Parallelization}

This is the phase, JPar also deviates from its predecessor. While the previous compiler would take two passes, one for generating dependencies between operations, and other for generating the tasks, the new version generates the dependencies when it is creating a task. This change allows to avoid creating tasks that would only be used for dependency purposes.

Firstly, this approach identifies parallelism with the finest granularity possible. Parallelizing all possible paths is not useful, as the overhead in scheduling may be significant. As such, the compiler decides to create tasks around parts of the code considered large enough. By default, only invocations and loops can be parallelized. In order to be parallelized, methods have to contain loops, at least 10 instructions, calls to other expensive methods, or recursive calls. The 10 instruction is a heuristic limit that can be configured.

The compiler performs three types of parallelization: Parallel Invocations, DO-ALL and DO-ACROSS. All of the three types can be nested inside each other.

\subsubsection{Parallel Invocation}

Every method invocation node is considered for parallelization if the target function is considered large enough. Invocations are converted into a Future call\cite{swaine2010back}, with a set of dependency tasks. Using Futures has then advantage of producing readable parallel Java, which the developer can use to learn, debug or to manually fine-tune.

The invocation is replaced by a call to the \verb|get()| method of the future. The original invocation is moved to a lambda that represents the computation (task body). That lambda is wrapped around a typed Future object that represents the asynchronous execution of the task. An example is shown in Listing~\ref{code:futures}, which shows a translation of code in Listing~\ref{code:signatures}, disregarding the fact that it is a recursive method. Recursive methods have the lambda converted into a static class, to avoid task creation overheads.

\vspace{15mm}

\lstset{language=Java,caption={The Fibonacci program translated with futures, without considering the special case.},label=code:futures}
\begin{lstlisting}
int f(int n) {
    if (RuntimeManager.shouldSeq())
        return jpar_sequential_version_of_f(n);

    if (n < 2) {
        return n;
    }
    Future<Integer> b_tmp = new Future<Integer>(task -> f(n-2));
    Future<Integer> a_tmp = new Future<Integer>(task -> f(n-1));
    int a = a_tmp.get();
    int b = b_tmp.get();
    return a + b;
}
\end{lstlisting}

When the Future object is instantiated, the task is marked for execution and an available thread may start to execute it. When the \verb|get()| method is called, the current task awaits for the execution of the task and reads its result.

The main decision to make is where to introduce the Future creation, maximizing parallelism while keeping the same semantics of the original program. The main requirements for the position of the Future creation are:

\begin{itemize}
    \item Must not be declared before the declaration of all used variables;
    \item Must not be declared before an expression which may return inside that method;
    \item Must not be declared before an expression which may change the control flow inside that block (break, continue);
    \item Must not be executed before an expression that may write to a variable accessed inside the lambda;
    \item Must not be executed before an expression that may reads a variable that is written inside the lambda;
    \item Must be before the Future \verb|get()| call.
\end{itemize}

Considering these requirements, Algorithm~\ref{algo:position} is used to find the best position to create the future, considering $\theta$ as the function that for an AST node returns its access permission set, $meth$ the method in which the invocation is found, $node$ the invocation being processed and $stmt$ the statement being analyzed. $block$ is the current block being analyzed. This block starts with the most outer scope (the method body) and moves deeper until it is the scope block in which the invocation is defined. This order tries to maximize how soon can the invocation start to execute. Finally, this algorithm outputs the Hard Dependency, which is the instruction after which the Future can be safely introduced, and the Soft Dependencies, a set of already defined tasks which the current future will have to wait to execute.

\begin{algorithm}
\begin{algorithmic}
\State $harddep \gets None$
\State $softdeps \gets \emptyset$
\For {$ stmt \in block $}
    \If {$control(meth) \in \theta(stmt) \vee control(block) \in \theta(stmt) $}
        \State $harddep \gets stmt $
        \State \textbf{continue}
    \EndIf
    \If {$\exists a, [read(a) \in \theta(stmt) \wedge write(a) \in \theta(node)] \vee [write(a) \in \theta(stmt) \wedge read(a) \in \theta(node)] \vee [write(a) \in \theta(stmt) \wedge write(a) \in \theta(node)]$}
        \If {$isTask(stmt)$}
            \State $softdeps \gets softdeps \cup stmt$
        \Else
            \State $harddep \gets stmt$
        \EndIf
    \EndIf
    \If {$stmt \supset node$}
        \State \textbf{break}
    \EndIf
\EndFor
\label{algo:position}
\caption{Algorithm to find the Hard Dependency and Soft dependencies for a Future for the current node}
\end{algorithmic}
\end{algorithm}

The invocation parallelization is completed when the Future declaration is at the right position, the Future call is at the invocation site, and there is a granularity control introduced in the beginning of each parallel method.

\subsubsection{Parallel DO-ALL}

Besides invocations, for and for-each loops are also targets for parallelization. However, there are two scenarios, DO-ALL and DO-ACROSS. First, we will focus on DO-ALL, when iterations of the loop are independent and have dependencies only with code outside the for loop.

In order to verify if this is the situation, accesses to arrays or arraylists are annotated with an indexed datagroup. This means that the code \lstinline|array[i] = 1| will have a permission $write(array[i])$ that is treated as a $write(array)$ for all code outside loops. Inside loops, the indexed permission is used to verify if reads and writes are independent. The verification performed is rather naïve, as it only considers for-loops in which the iteration variable is only increased or decreased. Nevertheless, it is possible to apply the polyhedral model, obtaining a better degree of parallelism.

The code generation of DO-ALL simply replaces the for loop with a call to a static helper method, that will dynamically create parallel tasks for slices of the range. The iteration code is defined as a lambda function passed to the helper method. The helper method will return a Future that can be used as a soft dependency for later Futures.

\subsubsection{Parallel DO-ACROSS}

In order to parallelize DO-ACROSS loops, the loop must contain the same conditions as for DO-ALL, but some write permissions are allowed inside the loop, namely operations that are commutative and associative. By default, the compiler considers for these tasks the operators \lstinline|+,-,*| and the methods \lstinline|Math.min(), Math.max()|. However, any over operation can be annotated as such, and will be parallelized using the same mechanism.

The compiler generates a Map-Reduce operation for the DO-ACROSS loop. The map lambda contains the loop code, saving the new changes inside the lambda, instead of the global state. The reduce lambda aggregates two states together. The return type of the operation is also a Future, in order to be used as a soft dependency.

\subsection{ÆminiumGPU Integration}

The ÆminiumGPU Compiler is a source-to-source compiler from Java-to-Java, in which the final Java code has some extra OpenCL code. The compiler targets lambda functions used inside Map-Reduce style of functions. For each of these lambdas, the compiler generates an OpenCL function. This function is compiled as a kernel during the compilation phase, but can also be dynamically compiled during execution, if merged with another function. This dynamic merging of functions into one kernel is used to avoid overheads in kernel scheduling and eventual memory transfers to and from the GPU.

It is important to notice that not all Java code can be translated to OpenCL. The ÆminiumGPU compiler does not support all method calls, non-local variables, for-each loops, object instantiation and exceptions. It does support a common subset between Java and C99 with some extra features like static accesses, calls to methods and references to fields of the Math object. 

Since DO-ALL and DO-ACROSS loops generate Map and Map-Reduce function calls, the lambdas can be automatically translated by the ÆminiumGPU compiler and handled by the ÆminiumGPU runtime, thus taking advantage of available GPU processing power.

\section{Runtime Execution}
\label{sec:runtime}

\subsection{Tasks and Dependencies}

The Æminium Runtime is a Java library that exposes APIs for expressing asynchronous execution of code. The Runtime is composed of modules that allow for an efficient execution of the source code, by leveraging the multiple hardware threads available.

The core concept of the Æminium Runtime is the task as a representation of code that can execute asynchronously. Tasks have a body, which can be represented as a lambda, an anonymous inner class or as a regular class (useful when doing recursive calls). Tasks are also defined by a set of dependencies on other tasks. If A depends on B, it means that A cannot execute before B is completed. Tasks can also have a parent task to represent subcomputations. If A is the parent of B, then A is only considered as completed when both the body of A has completed and B is considered as completed. Finally, tasks can be characterized with hints, such as Recursive, Loops, Small or Large. An example task graph can bee seen in Figure~\ref{fig:task_graph}, which represents 6 tasks with dependencies among them, as well as parent-child relationships.

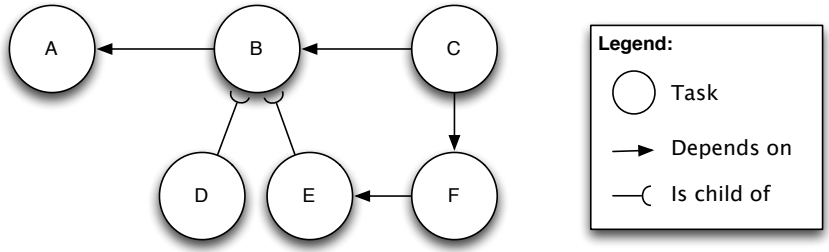
\botafig[0.6]{task_graph}{Example of a task graph.}

Each task can be of one of three types:

\begin{itemize}
	\item \textbf{Non-Blocking Tasks} are all operations that are purely computational.
	\item \textbf{Blocking Tasks} are tasks that have at least one operation that performs input or output, such as disk reads/writes, communication over sockets or other interactions with the Operative System.
	\item \textbf{Atomic Tasks} are tasks that cannot execute at the same time as other Atomic Tasks that share the same Data Group. The Data Group acts as the lock that each atomic task must acquire before executing and release after executing. However, two Atomic Tasks with different Data Groups can execute concurrently.
\end{itemize}

Figure~\ref{fig:runtime_phases} shows the lifecycle of tasks inside the Runtime. When a task is submitted to the runtime, along with its dependencies, the runtime analyzes if the dependencies are already met. If so, the task is sent to a queue for execution. If not, no action is performed at this point.

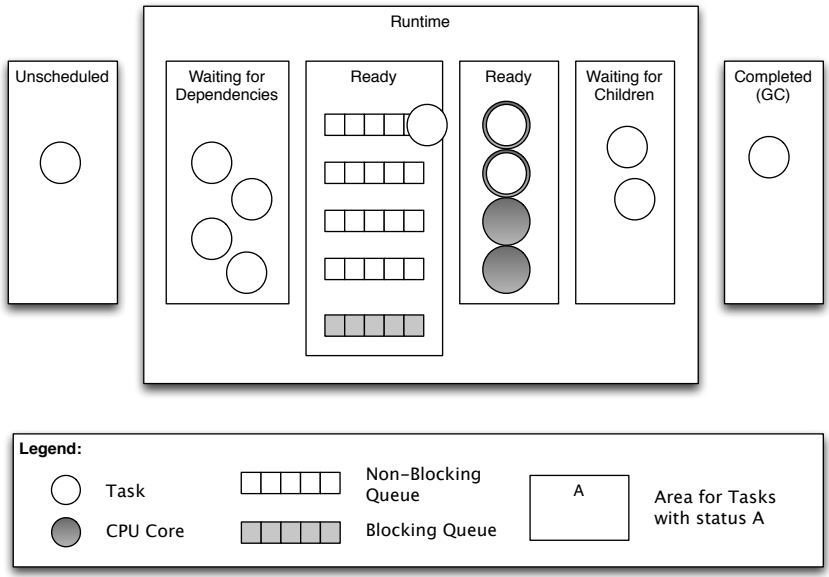
\botafig[0.5]{runtime_phases}{Runtime Areas for the different phases of the Task Lifecycle, for a quad-core machine.}

The ÆminiumRuntime does not create a Thread for each task, as the overhead would be very noticeable. Instead, there are always $n$ threads running, one for each processor core available in the machine (This number can be configurable per program execution). These threads are responsible for executing tasks that are considered ready. In order to reduce the locking on the queues, each thread has its own queue. When a thread finds its queue empty, it will ``steal'' a task from the queue of other thread. The ÆminiumRuntime has a few stealing algorithms, including a random steal, stealing from the largest queue or from the one with more dependent tasks.

Non-Blocking and Atomic tasks are stored on those regular queues. Since the execution of Blocking Tasks may take a long time to execute, because of OS dependencies such as Sockets or Files, they are added to a special ThreadPool-backed queue. This avoids blocking Work-Stealing workers with Blocking tasks.

When a task completes, it will check if there are any child tasks that were scheduled during execution and belong to the logical concern of the current task. When all child tasks have finished, the task is marked as completed, and it will notify both the dependent tasks and parent task that they do not need to wait for it anymore.

\subsection{Executing DO-ALL and DO-ACROSS}

Since each iteration may take a different time to execute, DO-ALL and DO-ACROSS loops cannot simply be divided in equal parts and executed in slices. In order to balance loads across cores, a more dynamic approach is required. We provide two different approaches: Binary Splitting and Lazy Binary Splitting\cite{tzannes2010lazy}. Binary Splitting divides the current range in two if the Decider module considers that it is still useful to create new tasks. If not, it executes the current range iterations immediately. With Lazy Binary Splitting, there is a parameter PPS which represents how frequently should we check if we should split the range in half. A PPS of 3 means that every 3 iterations the runtime checks if the remaining range should be split in two.

For DO-ACROSS loops, the Map-Reduce approach is better than creating lock-protected atomic blocks, since it avoids locking contention when all threads want to access that data. However, only associative and commutative operations can be converted into Map-Reduce. This is not a large problem, as most data-intensive computations are based on those operations, such as \lstinline|+,*, -|.

\subsection{Controlling Granularity}

One of the most important factors when executing irregular programs is to decide whether to execute a new task in parallel or inline the task body inside the current task. This decision has a great impact since the overhead in task creation is too high that can prevent recursive programs from having any speedups. The solution is to start calling the parallel version, but after a certain point, convert it to the sequential version of the method.

The Æminium Runtime provides several mechanisms for controlling the granularity of tasks:

\textbf{Maximum task recursion level (max-level)} - Divide-and-conquer algorithms create tasks in a tree-shaped structure. In order to avoid the creation of too many tasks, the cut-off limit may be defined by the depth of the recursion\cite{duran2008evaluation}, which can be calculated by the number of ancestors of the running task.

This approach limits the depth of the task hierarchy graph, which may be suitable for more balanced parallelism programs, but not for more dynamic unbalanced programs.

\textbf{Maximum number of tasks (max-tasks)} - In this approach, tasks are created until the total number of active tasks reaches a certain threshold\cite{duran2008evaluation}. After that point, all new computations are inlined instead of spawning another thread. When the number of active tasks lowers, new tasks can be created until the threshold is reached again.

The threshold in this approach is typically defined as the number of processor threads on the machine, adapting to different machines, but being oblivious to other factors such as memory and processor speed.

\textbf{Load Based} - This simple heuristic is based on whether all cores are being used or not. A new task is only created if there is at least one idle core\cite{duran2008adaptive}.

\textbf{Surplus Queued Task Count} - This approach is included in Java's Fork Join framework\cite{lea2000java} and it relies on the size of work-stealing queues. Before creating a new task, the number of queued tasks in the current thread that exceeds the number of tasks in other queues is compared to a threshold limit (usually 3 in existing ForkJoin benchmarks).

In order to decrease the overhead of computing the size of queues, the size of other queues is estimated from the size of the current queue after applying a factor of (number of idle threads $/$ active threads), because idle threads are known to have 0 tasks in their queue. This estimation assumes a regular distribution among threads, which may not always happen.

\textbf{Adaptive Tasks Cut-Off (ATC)} - Adaptive Tasks Cut-Off\cite{duran2008adaptive} changes the policy of the cut-off mechanisms according to the recursion lifecycle. Tasks are only created if two conditions are met. The first is that there are fewer tasks than the number of threads on a given recursion level. This condition forces the threads to expand in depth, creating work for all threads and being within a certain bound limit. The second condition is that the depth-level is less than a certain threshold. This is, in fact, the usage of \emph{max-level} and \emph{max-tasks} together.

ATC adds a profiler that saves information regarding how much time a sub-tree takes to execute, and predicts further subtrees (if the prediction is larger than 1ms, the task will not be created). This is, however, based on the assumption that all tasks inside a level have a similar behavior, which does not happen in unbalanced parallelism.

\textbf{Maximum Queue Size} - We introduce this new approach, which limits the number of tasks in the local queue to a certain threshold This approach differs from \emph{maxtasks} in only looking at the local queue, instead of all the queues, reducing the time by not accessing information from other threads. If the threshold is one or two tasks higher than the threshold of \emph{max-tasks}, queues will have extra tasks that can be stolen by other threads.

\textbf{Stack Size} - In recursive divide-and-conquer programs, the recursion limit of the platform imposes heavy limitations on the parallelization of programs. Recursive calls are necessary to inline the execution of tasks inside the same worker thread. As such, the performance of programs decreases when the stack grows beyond a certain size.

\textbf{System Monitoring} - Instead of looking to the task and runtime state, this decider method analyses the system load. If it is below a given CPU occupation and below a given memory occupation, then a new task is created. Both the CPU and memory occupations can be configurable.

The Runtime can use any of these methods for each program execution. It is also possible to combine two or more of the methods at the cost of increasing the overhead of the decision.

\subsection{ÆminiumGPU Integration}

The ÆminiumGPU Runtime allows the execution of data-parallel operations (such as map and reduce) on GPUs. Operations are lazy and are only executed when the result is needed. This differs from regular Java semantics, as it can avoid unnecessary overheads in the case two GPU operations are chained. In that case, we merge the GPU kernels into one, making only one data copy to each side, and starting only one kernel.

The first step is to decide how to divide work between the GPU and the CPU cores. If the GPU is going to be used, a OpenCL kernel is compiled, data is sent to the GPU and the kernel executes. After completion, data is copied back to the main memory.

The decision whether to use the GPU is firstly done by analyzing the size of input data. If it is below a given threshold (which depends on the machine) and the operation is complex enough, then it is executed on the regular Æminium Runtime as a DO-ALL or DO-ACROSS loop. If it is complex enough, it can execute on the GPU or in a mix of GPU and CPU. This decision has been improved by using Machine Learning\cite{fonseca2013aeminiumgpu}, in which features from static analysis (number of memory writes and reads, global vs local accesses, cyclomatic complexity, number of branch instructions and number of mathematical intensive operations) and runtime information (size of data, number of chained operations) are used to decide which platform to use for execution.

\section{Evaluation}
\label{sec:evaluation}

In this section, we will evaluate several design options of the JPar Compiler, as well as Runtime configuration for its programs. We then compare it with OoOJava and with a manually parallelized version. The experiments shown were performed on a 12-core machine with a Intel Xeon X5660 @ 2.8GHz processor with 12 cores and 24 hyper-threads, and 24GB of RAM. The machine was running Ubuntu 14.04 server 64-Bit and Java Hotspot 64-Bit Server JVM. The machine was chosen for it's high number of cores. Unless specified, programs were executed 7 times and the average value was used.

\subsection{Benchmark Programs}

In order to evaluate the performance of the compiler, we used the sequential version of 8 programs from the Æminium Benchmark Suite\cite{aebenchmarks}. The configuration for each program is described in Table~\ref{table:benchmark}, as well as the parallelization performed for each program.

\botatab{| p{3cm} | l | l | l | p{3cm} | l |}{benchmark}{Description of the programs used in the benchmark}{
\hline
Program & Parallelism & Input size \\ \hline
Black-Scholes & DO-ACROSS & 100000 \\ \hline
FFT & Recursive & 16777216\\ \hline
Fibonacci & Recursive & n=51\\ \hline
Health & Recursive, DO-ALL & n=6\\ \hline
Integrate & Recursive & s=-2101, e=1700, error=$10^{-14}$\\ \hline
MergeSort & Recursive & n=251658240 \\ \hline
N-Body & DO-ALL & it=10, bodies=25000 \\ \hline
Pi & DO-ACROSS & n=1500000000 \\ \hline
}

\subsection{Compiler Granularity Control}

The JPar compiler performs a selection of whether tasks should be parallelizable, based on the number of instructions. To evaluate whether this is useful, we compared two versions of the compiler, one with fine-grained parallelism, and other that only creates tasks if there are 10 or more instructions in the body of the task, or have recursive calls or loops.

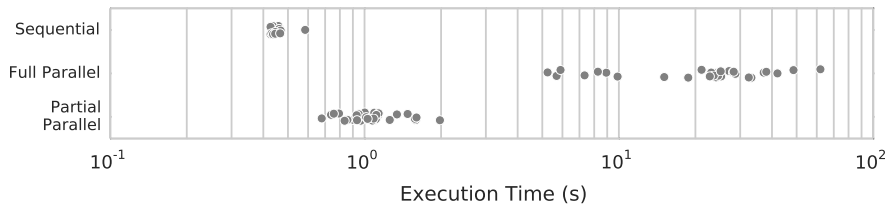
\botafig[0.85]{compiler_granularity_control}{Distribution of execution times for each version of the FFT program of a random array with 262144 elements.}

We used the FFT benchmark because it works with an array of Complex objects, each with several lightweight methods, such as sin, cos, tan, times and divide. The Full Parallel version converts invocations that can occur in parallel into tasks, while the Partial parallel will only create parallel tasks for the main recursive function. Figure~\ref{fig:compiler_granularity_control} shows the distribution of execution times of both versions, as well as of the sequential version. With an input size of 262144 elements, there is no speedup in any of the versions. However, with higher input sizes, the Full Parallel version will have take more than one hour to execute, while the Partial Parallel version would provide speedups. Thus, we confirm that enabling a threshold for task parallelization benefits programs.

\subsection{Binary vs Lazy Binary Splitting}

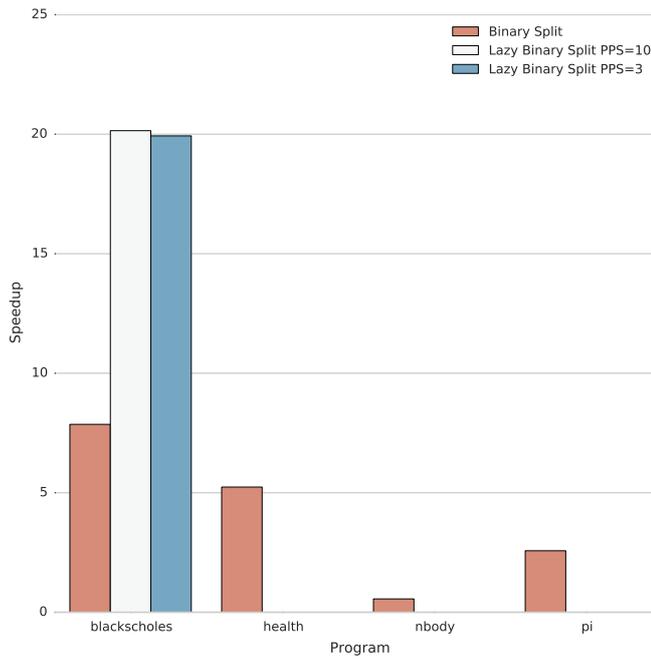
\botafigaqui[0.6]{binary_speedup}{Average speedup of Binary Split and Lazy Binary Split (PPS=3 and 10) versions of the programs with loops.}

Programs that have for loops parallelized, can generate tasks in two different ways: Binary Splitting or Lazy Binary Splitting. For Lazy Binary Splitting, we used the recommended value of 3 for the PPS parameter, and a higher value of 10 for less frequent decisions. Figure~\ref{fig:binary_speedup} shows the speed-up over sequential programs of the three approaches in program with loops. The Lazy Binary Split version achieved best results in the Black-Scholes program, running in less than half of the time as its Binary Split counterpart, but could not complete the other programs within a 5-minute time-out, resulting in no speed-up. The conclusion is that Binary Split is a conservative approach that can be used for any program, while Lazy Binary Split can be used to achieve best results, but needs to be applied after verifying there is a speedup in that particular program.

\subsection{Cutoff Mechanism}

One of the most important variables for tuning a parallel program is the granularity of tasks. Besides the compilation-time decision of parallelizing a task or not, the Æminium Runtime has several automatic granularity control mechanisms. Figure~\ref{fig:cutoff} shows the speed-up achieved by each method on the 8 programs. For each mechanism, several parameters were previously tested, and we used the best.

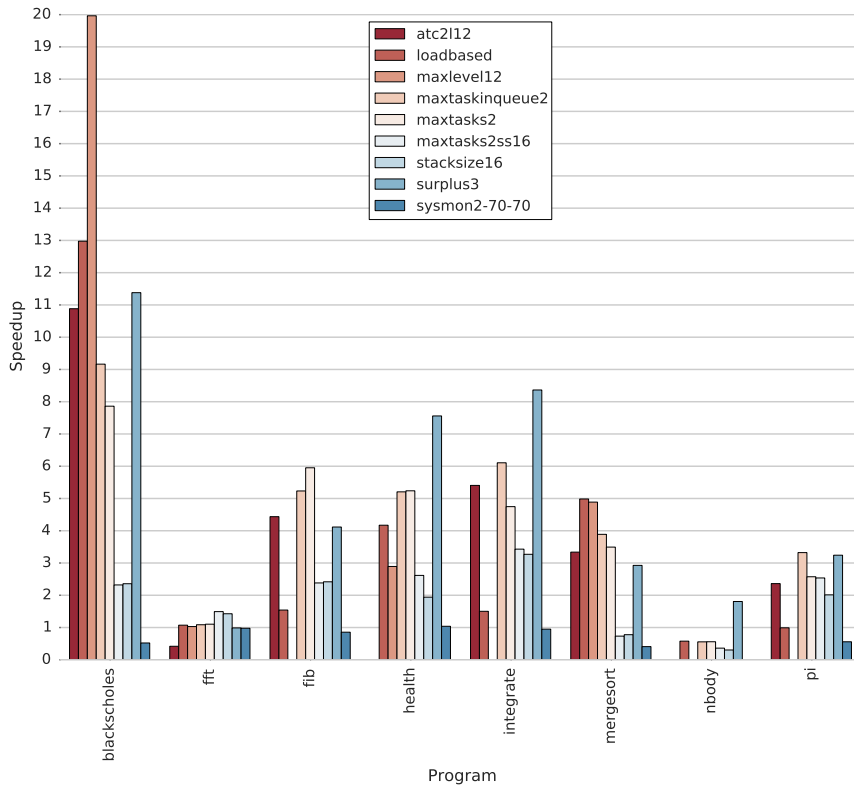
\botafig[0.55]{cutoff}{Speed-ups of different cut-off approaches for each program.}

Black-Scholes is a program with a high number of tasks and each task performs very small work. Creating extra-tasks in this kind of program brings a high penalty, because the task overhead is several times more expensive than executing the task itself. \emph{max-level} is the mechanism with the best performance since all loop iterations take equal time to execute. However, a maximum level of other value than 12 would not perform as well. The same mechanism does not perform well on Fibonacci or Integrate given the very irregular and unbalanced nature of the problem. The skewness in the computation tree would leave one task with a time-consuming task similar that of the sequential version.

Black-Scholes is the program that takes more advantage of being parallelized on this machine. ATC, Load Balance, Max-Level and Surplus show a good performance. When cutting parallelism by the stack depth or system resources, too many tasks are created, and there is an high unnecessary overhead.

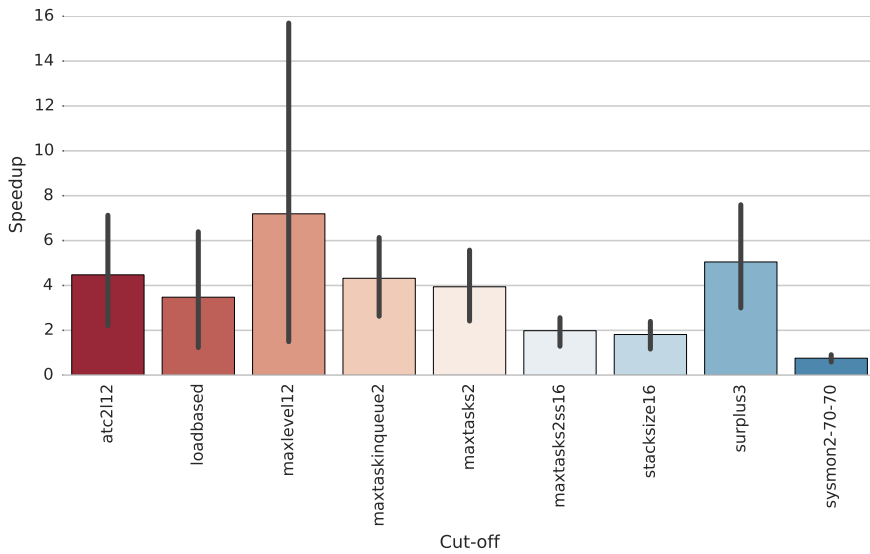
\botafig[0.6]{cutoff_average}{Average speed-ups of different cut-off approaches across the 8 different programs.}

The FFT program did not achieve high speedup values because of memory issues. Each recursive call would allocate a large array of memory, and this revealed to be a performance hit on the java platform. The approaches that limited the JVM stack depth achieved the best results, but still under the desired values. The same happened with N-Body simulation because of compiler-time granularity. N-Body is made of two loops, and the JPar compiler over-parallelizes the program by trying to convert both loops in DO-ALL operations. Since the inner loop is small enough to be considered only one task, the compiler introduces a high overhead by trying to parallelize a block that is already small enough. 

The comparison between approaches for the overall benchmark can be seen in Figure~\ref{fig:cutoff_average}. Despite Max-Level having the highest speedup, we used as default the Surplus3 approach because of its lower variance. Most of the difference of speedup of Max Level is due to only one program: Black-Scholes.

\subsection{Comparison with Other Approaches}

We evaluated the JPar compiler against two other approaches: OoOJava and a human programmer. The source programs were annotated with the \emph{sese} statements to identify main parallel tasks. Without these annotations, the generated version would be equivalent to the sequential version. Since the OoOJava compiler generates C code, a sequential version using the same compiler is also presented. The OoOJava compiler was unable to compile Black-Scholes, Health and the N-Body programs, failing to identify dependencies during code-generation. The Æminium Benchmark includes versions of the programs manually written on top of the Fork-Join framework, and on top of the Æminium Runtime. Manual or automatic cut-off mechanisms were selected using a manual local search across mechanisms and parameters.

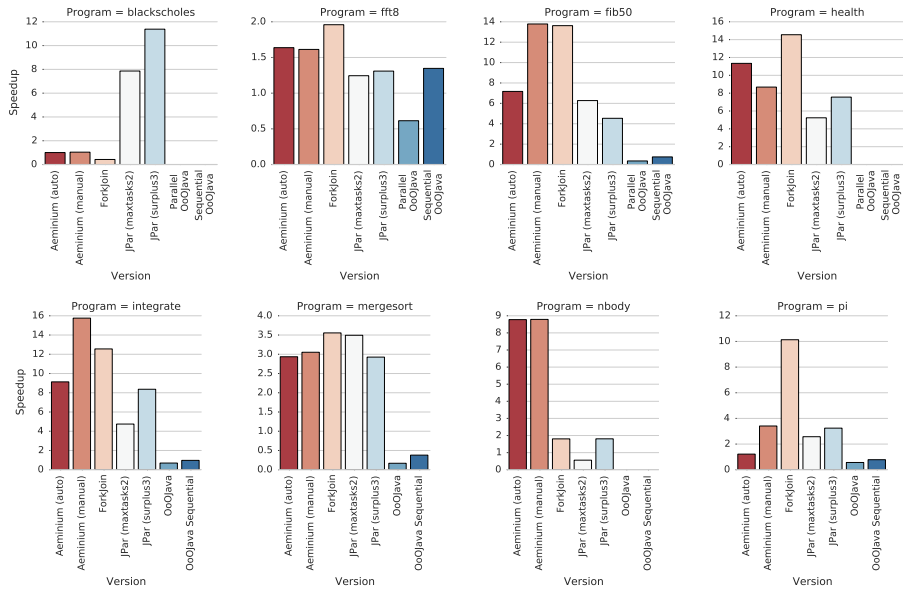
\botafigdouble[0.45]{ooo_comparison_0}{ooo_comparison_1}{Speed-up comparison between JParCompiler, Human and OoOJava approaches. The JPar version includes the two best performant cut-off mechanisms, maxtasks2 and surplus3. The Human version includes one version on top of the Fork-Join framework, and two on top of the Æminium Runtime, one with a manual cut-off and other with an automatic cut-off mechanism. For the OoOJava compiler, the serial and parallel versions are shown.}

Figure~\ref{fig:ooo_comparison_0} shows the speedup achieved by the three approaches, each with different configurations of platform, granularity control mechanism and, in the case of OoOJava, the sequential version is also included. Overall, the performance of JPar compiler was superior to that of OoOJava. OoOJava did not even achieve a speedup compared with the OoOJava sequential version. The reason is that the automatic granularity control mechanism of the OoOJava over-parallelizes, resulting in overheads on smaller tasks and on a heavier contention in locking. The results presented in \cite{jenista2011ooojava} mask the overhead by introducing a manual cut-off decision in the sequential programs, which should not be aware that they are parallel.

Overall, the human versions Fork-Join and Æminium programs performed better than those of JParCompiler, which is expected, since the programmers have a better knowledge of the nature of the program, and select the best parts to parallelize. Black-Scholes is an exception, since the Human version only parallelizes inside each loop, and does not consider parallelizing different tasks at the same time. This is a case where the compiler could find parallelism hidden in plain sight, and this can be used to further improve the benchmarks. In the other cases, the Human versions outperformed the JPar compiler with the best general cut-off. It is important to notice that the best threshold in some programs would show a speed-up similar to the human version, however we decided not to include a manual tuning of the parallel version generated by the JPar compiler.

\section{Conclusion}
\label{sec:conclusion}

We have presented the JPar compiler and the Æminium framework for automatically parallelizing sequential programs. By analyzing the data dependencies in the sequential program, we were able to conservatively extract parallelism without changing the program semantics. We have improved over our previous work by performing the data-dependency analysis in one pass, and by generating source-code similar to the original, but with futures replacing parallel computations. This change allows developers to understand how parallelization occurs, and how to improve it.

We have also studied two granularity control mechanisms. The JPar compiler only considers for parallelization tasks that are considered large enough. This change introduced speedups in several programs of the benchmark, that would not have it otherwise. We have also applied several existing runtime cut-off mechanisms, as well as three new (StackSize, System Monitor, and a combination of StackSize with Max Tasks) that can be used to improve the performance of programs. We have also studied the usage of Binary versus Lazy Binary Splitting.

Finally, we compared our compiler with another state-of-the-art compiler, OoOJava, and with human parallelization. While our results were not as good as if a human would write and fine-tune the programs, JPar generated programs outperformed OoOJava.

For future work, we intend to select the best granularity control mechanism for a given program by looking at its structure. This can be performed using machine learning techniques over a large dataset of programs. A low-overhead combination of mechanisms that can be used to improve any program is also currently being explored. Another aspect which needs improvement is to better access the granularity of tasks at compile-time, for which a cost model can be of use.

\begin{acknowledgements}
This work would not have been possible without the contributions to the Æminium language and runtime from Sven Stork, Paulo Marques and Jonathan Aldrich.
\end{acknowledgements}

\bibliographystyle{spmpsci}      % mathematics and physical sciences
\bibliography{bibliography}

\begin{thebibliography}{10}
\providecommand{\url}[1]{{#1}}
\providecommand{\urlprefix}{URL }
\expandafter\ifx\csname urlstyle\endcsname\relax
  \providecommand{\doi}[1]{DOI~\discretionary{}{}{}#1}\else
  \providecommand{\doi}{DOI~\discretionary{}{}{}\begingroup
  \urlstyle{rm}\Url}\fi

\bibitem{amini2012par4all}
Amini, M., Creusillet, B., Even, S., Keryell, R., Goubier, O., Guelton, S.,
  McMahon, J.O., Pasquier, F.X., P{\'e}an, G., Villalon, P.: Par4all: From
  convex array regions to heterogeneous computing.
\newblock In: IMPACT 2012: Second International Workshop on Polyhedral
  Compilation Techniques HiPEAC 2012 (2012)

\bibitem{ayguade2009design}
Ayguad{\'e}, E., Copty, N., Duran, A., Hoeflinger, J., Lin, Y., Massaioli, F.,
  Teruel, X., Unnikrishnan, P., Zhang, G.: The design of openmp tasks.
\newblock Parallel and Distributed Systems, IEEE Transactions on
  \textbf{20}(3), 404--418 (2009)

\bibitem{ayguade2008experimental}
Ayguad{\'e}, E., Duran, A., Hoeflinger, J., Massaioli, F., Teruel, X.: An
  experimental evaluation of the new openmp tasking model.
\newblock In: Languages and Compilers for Parallel Computing, pp. 63--77.
  Springer (2008)

\bibitem{banerjee1993automatic}
Banerjee, U., Eigenmann, R., Nicolau, A., Padua, D.A., et~al.: Automatic
  program parallelization.
\newblock Proceedings of the IEEE \textbf{81}(2), 211--243 (1993)

\bibitem{bik1997automatically}
Bik, A.J., Gannon, D.B.: Automatically exploiting implicit parallelism in java.
\newblock Concurrency - Practice and Experience \textbf{9}(6), 579--619 (1997)

\bibitem{bondhugula2008automatic}
Bondhugula, U., Baskaran, M., Krishnamoorthy, S., Ramanujam, J., Rountev, A.,
  Sadayappan, P.: Automatic transformations for communication-minimized
  parallelization and locality optimization in the polyhedral model.
\newblock In: Compiler Construction, pp. 132--146. Springer (2008)

\bibitem{chamberlain2007parallel}
Chamberlain, B., Callahan, D., Zima, H.: Parallel programmability and the
  chapel language.
\newblock International Journal of High Performance Computing Applications
  \textbf{21}(3), 291--312 (2007)

\bibitem{chan2004run}
Chan, B., Abdelrahman, T.S.: Run-time support for the automatic parallelization
  of java programs.
\newblock The Journal of Supercomputing \textbf{28}(1), 91--117 (2004)

\bibitem{charles2005x10}
Charles, P., Grothoff, C., Saraswat, V., Donawa, C., Kielstra, A., Ebcioglu,
  K., Von~Praun, C., Sarkar, V.: X10: an object-oriented approach to
  non-uniform cluster computing.
\newblock In: ACM SIGPLAN Notices, vol.~40, pp. 519--538. ACM (2005)

\bibitem{chen2003jrpm}
Chen, M.K., Olukotun, K.: The jrpm system for dynamically parallelizing java
  programs.
\newblock In: Computer Architecture, 2003. Proceedings. 30th Annual
  International Symposium on, pp. 434--445. IEEE (2003)

\bibitem{dagum1998openmp}
Dagum, L., Enon, R.: Openmp: an industry standard api for shared-memory
  programming.
\newblock Computational Science \& Engineering, IEEE \textbf{5}(1), 46--55
  (1998)

\bibitem{dave2009cetus}
Dave, C., Bae, H., Min, S.J., Lee, S., Eigenmann, R., Midkiff, S.: Cetus: A
  source-to-source compiler infrastructure for multicores.
\newblock Computer (12), 36--42 (2009)

\bibitem{dominguez2013evaluating}
Dominguez, R.M.: Evaluating different java bindings for opencl  (2013)

\bibitem{duran2008adaptive}
Duran, A., Corbal{\'a}n, J., Ayguad{\'e}, E.: An adaptive cut-off for task
  parallelism.
\newblock In: High Performance Computing, Networking, Storage and Analysis,
  2008. SC 2008. International Conference for, pp. 1--11. IEEE (2008)

\bibitem{duran2008evaluation}
Duran, A., Corbal{\'a}n, J., Ayguad{\'e}, E.: Evaluation of openmp task
  scheduling strategies.
\newblock In: OpenMP in a new era of parallelism, pp. 100--110. Springer (2008)

\bibitem{feautrier1996automatic}
Feautrier, P.: Automatic parallelization in the polytope model.
\newblock In: The Data Parallel Programming Model, pp. 79--103. Springer (1996)

\bibitem{aebenchmarks}
Fonseca, A.: {Æminium Benchmark Suite}.
\newblock \url{https://github.com/AEminium/AeminiumBenchmarks} (2013).
\newblock [Online; accessed 23-October-2013]

\bibitem{fonseca2013aeminiumgpu}
Fonseca, A., Cabral, B.: {\AE}miniumgpu: An intelligent framework for gpu
  programming.
\newblock In: Facing the Multicore-Challenge III, pp. 96--107. Springer (2013)

\bibitem{frigo1998implementation}
Frigo, M., Leiserson, C.E., Randall, K.H.: The implementation of the cilk-5
  multithreaded language.
\newblock In: ACM Sigplan Notices, vol.~33, pp. 212--223. ACM (1998)

\bibitem{hogen1992automatic}
Hogen, G., Kindler, A., Loogen, R.: Automatic parallelization of lazy
  functional programs.
\newblock In: ESOP'92, pp. 254--268. Springer (1992)

\bibitem{jenista2011ooojava}
Jenista, J.C., Demsky, B.C., et~al.: Ooojava: software out-of-order execution.
\newblock In: ACM SIGPLAN Notices, vol.~46, pp. 57--68. ACM (2011)

\bibitem{lea2000java}
Lea, D.: A java fork/join framework.
\newblock In: Proceedings of the ACM 2000 conference on Java Grande, pp.
  36--43. ACM (2000)

\bibitem{lee2009openmp}
Lee, S., Min, S.J., Eigenmann, R.: Openmp to gpgpu: a compiler framework for
  automatic translation and optimization.
\newblock ACM Sigplan Notices \textbf{44}(4), 101--110 (2009)

\bibitem{leino2002using}
Leino, K., Poetzsch-Heffter, A., Zhou, Y.: Using data groups to specify and
  check side effects.
\newblock ACM SIGPLAN Notices \textbf{37}(5), 246--257 (2002)

\bibitem{marlow2009runtime}
Marlow, S., Peyton~Jones, S., Singh, S.: Runtime support for multicore haskell.
\newblock In: ACM Sigplan Notices, vol.~44, pp. 65--78. ACM (2009)

\bibitem{spoon}
Pawlak, R., Monperrus, M., Petitprez, N., Noguera, C., Seinturier, L.: Spoon
  v2: Large scale source code analysis and transformation for java.
\newblock Tech. Rep. hal-01078532, INRIA (2006).
\newblock \urlprefix\url{https://hal.inria.fr/hal-01078532}

\bibitem{rafael2014dependency}
Rafael, J., Correia, I., Fonseca, A., Cabral, B.: Dependency-based automatic
  parallelization of java applications.
\newblock In: Euro-Par 2014: Parallel Processing Workshops, pp. 182--193.
  Springer (2014)

\bibitem{senghor2013fjcomp}
Senghor, A., Konate, K.: Fjcomp, a java parallelizing compiler for dealing with
  divide-and-conquer algorithm.
\newblock In: Computer Applications Technology (ICCAT), 2013 International
  Conference on, pp. 1--5. IEEE (2013)

\bibitem{steele2006parallel}
Steele, G.: Parallel programming and parallel abstractions in fortress.
\newblock Lecture Notes in Computer Science \textbf{3945}, 1 (2006)

\bibitem{stork2014aeminium}
Stork, S., Naden, K., Sunshine, J., Mohr, M., Fonseca, A., Marques, P.,
  Aldrich, J.: {\AE}minium: A permission-based concurrent-by-default
  programming language approach.
\newblock ACM Transactions on Programming Languages and Systems (TOPLAS)
  \textbf{36}(1), 2 (2014)

\bibitem{swaine2010back}
Swaine, J., Tew, K., Dinda, P., Findler, R.B., Flatt, M.: Back to the futures:
  incremental parallelization of existing sequential runtime systems.
\newblock In: ACM Sigplan Notices, vol.~45, pp. 583--597. ACM (2010)

\bibitem{tzannes2010lazy}
Tzannes, A., Caragea, G.C., Barua, R., Vishkin, U.: Lazy binary-splitting: a
  run-time adaptive work-stealing scheduler.
\newblock ACM Sigplan Notices \textbf{45}(5), 179--190 (2010)

\bibitem{wang2013mp}
Wang, C., Li, X., Zhang, J., Zhou, X., Nie, X.: Mp-tomasulo: A dependency-aware
  automatic parallel execution engine for sequential programs.
\newblock ACM Transactions on Architecture and Code Optimization (TACO)
  \textbf{10}(2), 9 (2013)

\bibitem{zhao2005loop}
Zhao, J., Rogers, I., Kirkham, C., Watson, I.: Loop parallelisation for the
  jikes rvm.
\newblock In: Parallel and Distributed Computing, Applications and
  Technologies, 2005. PDCAT 2005. Sixth International Conference on, pp.
  35--39. IEEE (2005)

\end{thebibliography}

\end{document}